\documentstyle[12pt]{article}
\newcommand{\p}[1]{(\ref{#1})}\newcommand{\bi}[1]{\bibitem{#1}}

\newcommand{\nn}{\nonumber\\}

\begin{document}

\renewcommand{\thefootnote}{\fnsymbol{footnote}}
\thispagestyle{empty}

\begin{center}
{\Large \bf Nonlinear realization of superconformal symmetry and
Liouville equation superextensions \footnote{Talk given at the International
Workshop "Supersymmetry and Quantum Symmetries", JINR, Dubna, Russia,
July 26-31, 1999}}
\vskip 0.5cm

\vskip 0.5cm

{\bf A. A. Kapustnikov \footnote{E-mail: kpstnkv@creator.dp.ua}}

{\it Department of Physics, Dnepropetrovsk University, \\
     320625, Dnepropetrovsk, Ukraine}

\vskip 1.5cm

{\bf Abstract}
\end{center}

It is shown that the method of nonlinear realization of local
supersymmetry being applied to the $n=(1,1)$ superconformal symmetry allows
one reduce the new version of the super-Liouville equation to the ordinary
one owing to the relaxation of the auxiliary equation of motion fixing the
gauge parameters.

\setcounter{page}1
\renewcommand{\thefootnote}{\arabic{footnote}}\setcounter{footnote}0

\section{Introduction}

It was reviled in \cite{biku} that the general solution of the string-
inspired
nonlinear equations describing the intrinsic geometry of the bosonic string
worldsheet in the geometrical approach \cite{geom,bpstv,b}
can be
represented  in terms of the two sets of left- and right-moving Lorentz
harmonic variables. The latter are defined as the coordinates of compact
coset space isomorphic to the $D$-dimensional sphere
\begin{equation}\label{cs}
  S_{D-2}={SO(1,D-1)
\over SO(1,1) \times SO(D-2) \times K_{D-2}},
\end{equation}
with the Borel subgroup of Lorentz group in the denominator of the
fraction \p{cs}. A natural way of extending these results on the case of
superstring is to search for the corresponding solutions of supersymmetric
equations of motion describing the embedding of superstrings into the flat
target superspaces. The simplest example of such a pattern of equations
inherent to the $N=2,~D=3$ superstring
is the new version of the $n=(1,1)$ super-Liouville equation \cite{bsv}
\footnote{The old version
given in \cite{ck,ikr} is also acceptable.}
\begin{equation}\label{n11sl}
D_-D_+W = e^{2W}\Psi^+_L\Psi^-_R,~~~
D_{\pm} = \partial_{\pm} + i\eta^{\pm}\partial_{\pm\pm},
\end{equation}
\begin{eqnarray}\label{gfe}
D_+\Psi^+_L - 2(D_+W)\Psi^+_L = 1, \\
D_-\Psi^-_R - 2(D_-W)\Psi^-_R = 1, \nonumber
\end{eqnarray}
where
\begin{equation}\label{W}
W(\xi^{\pm\pm}, \eta^{\pm}) = u(\xi^{\pm\pm}) + i\eta^+\psi^-(\xi^{\pm\pm})
+ i\eta^-\psi^+(\xi^{\pm\pm}) + i\eta^-\eta^+F(\xi^{\pm\pm}),
\end{equation}
\begin{eqnarray}\label{Psi}
\Psi^+_L(\xi^{++}, \eta^+) &=& \omega^+(\xi^{++}) + \eta^+F_L(\xi^{++}), \\
\Psi^-_R(\xi^{--}, \eta^-) &=& \omega^-(\xi^{--}) + \eta^-F_R(\xi^{--}),   \nonumber
\end{eqnarray}
are the worldsheet superfields.

In this report we would like to show that as well as in the bosonic case Eqs.
\p{n11sl}, \p{gfe} can be solved exactly in terms of the Lorentz harmonics
variables parametrizing the coset space \p{cs} but unlike to them valued on
the worldsheet superspace
${\bf R}^{(2 \mid 2)}~ =~ \{\xi^{++}, \eta^+; \xi^{--}, \eta^-\}$
and restricted by the special covariant constraints deriving from the
nonlinear realization of the superconformal symmetry.

\section{Gauge and superconformal symmetries}
\subsection{Linear realization}
It is not hard to verify that the flat spinor covariant derivatives entering
the Eqs. \p{n11sl}, \p{gfe} are transformed homogeneously
\begin{equation}\label{trD}
D^{\prime}_{\pm} = (D_{\pm}\eta^{\pm\prime})^{-1}D_{\pm}.
\end{equation}
with respect to the two copies of one dimensional superconformal
transformations
\begin{eqnarray}\label{sct}
\xi^{\pm\pm\prime} &=& \xi^{\pm\pm} + a^{\pm\pm} +
i\eta^{\pm} \epsilon^{\pm}\sqrt{1 +
a^{\pm\pm\prime}},~~~a^{\pm\pm\prime} = \partial_{\pm\pm}a^{\pm\pm}  \\
\eta^{\pm\prime} &=& \epsilon^{\pm} + \eta^{\pm}\sqrt{1 +  a^{\pm\pm\prime} +
i\epsilon^{\pm}\epsilon^{\pm\prime}}, ~~~~\epsilon^{\pm\prime} =
\partial_{\pm\pm}\epsilon^{\pm},
\nonumber
\end{eqnarray}
restricted by the condition
\begin{equation}\label{Del}
D_{\pm}\xi^{\pm\pm\prime} - i\eta^{\pm}D_{\pm}\eta^{\pm\prime} = 0.
\end{equation}
This indicate that the following gauge transformations of superfields
\begin{eqnarray}\label{gtSF}
W^{\prime}(\xi^{\pm\pm\prime}, \eta^{\pm\prime}) &=&
W(\xi^{\pm\pm}, \eta^{\pm})
 - \frac{1}{4}ln(D_+\eta^{+\prime}) -
\frac{1}{4}ln(D_-\eta^{-\prime}), \\
\Psi^{+\prime}_L(\xi^{++\prime}, \eta^{+\prime}) &=&
(D_+\eta^{+\prime})^{-1/2}\Psi^+_L(\xi^{++}, \eta^+), \nn
\Psi^{-\prime}_R(\xi^{--\prime}, \eta^{-\prime}) &=&
(D_-\eta^{-\prime})^{-1/2}\Psi^-_R(\xi^{--}, \eta^-),
\nonumber
\end{eqnarray}
leaves intact the form of the Eq. \p{n11sl}. At the same time the second
Eq. \p{gfe} is not changed as well only when gauge is completely fixed
\begin{equation}\label{gfc}
(D_-\eta^{-\prime})^{-3/2} = 1.
\end{equation}
It was shown in \cite{bsv} that this gauge condition impose very essential
restrictions on the superfields \p{W}, \p{Psi} removing all their components
excepting leading once $u(\xi^{\pm\pm})$ and $\omega^{\pm}(\xi^{\pm\pm})$.
The simplest way of achieving this result is to transit to the nonlinear
realization of superconformal symmetry in which the gauge fixing Eq. \p{gfe}
can be initially something relaxed and then exactly solved.
In the next Section we are going to construct this
realization following closely to Ref. \cite{ik}.

\subsection{Nonlinear realization}
 Let us suppose that the v.e.v. of the component fields
$F_L(\xi^{++})$ and $F_R(\xi^{--})$ in \p{Psi} are not
equal to zero and as consequence of this the local supersymmetry \p{sct} is
actually spontaneously broken. In this case the fermionic components
$\omega^{\pm}(\xi^{\pm\pm})$ acquire the meaning
of the corresponding Goldstone fermions and one can exploit them
for a singling out of the ordinary Liouville equation from the system
\p{n11sl}, \p{gfe} in a manifestly covariant manner. Indeed, it is
well-known that
in the models with spontaneously broken supersymmetry all the SFs
becomes reducible \cite{eiak}, \cite{ik}. Their irreducible parts are
transformed, however, universally with respect to the action of the original
supergroups, as the linear representations of the underlying
unbroken subgroups but with the parameters depending nonlinearly on the
Goldstone fermions. There is always the possibility to impose on these SFs
some absolutely covariant restrictions providing to remove
out from them undesirable degrees of
freedom. Here we can take advantage of possibilities of this approach for
deriving the relevant solution of the Eqs. \p{n11sl}, \p{gfe}.

For the beginning let us consider some special aspects of the
nonlinear realization of superconformal symmetry in superspace. As was shown
in Refs. \cite{cwz}, \cite{ik} for this purpose
we need firstly splits the general finite element of the group \p{sct}
\begin{equation}\label{G}
G(\zeta) \equiv \zeta^{\prime},
\end{equation}
where $\zeta = \{\xi^{\pm\pm}, \eta^{\pm}\}$, onto the product of elements
of two successive transformations
\begin{equation}\label{suctr}
G(\zeta) = K(G_0(\zeta)).
\end{equation}
In Eq. \p{suctr} the following standard notations are used. As before the
$G_0(\zeta)$ denotes the "primes" coordinates $\zeta^{\prime}$ but index
{\it zero} means that they referring now only to the stability subgroup
\begin{eqnarray}\label{Gvac}
\xi^{\pm\pm\prime} &=& \xi^{\pm\pm} + a^{\pm\pm}(\xi^{\pm\pm}), \\
\eta^{\pm\prime} &=& \eta^{\pm}\sqrt{1 + \partial_{\pm\pm}a^{\pm\pm}}.
\nonumber
\end{eqnarray}
By the definition we suppose also that the stability subgroup includes only
the ordinary conformal transformations (parameters
$a^{\pm\pm}(\xi^{\pm\pm})$) of the bosonic coordinates $\xi^{\pm\pm}$ and the special
scale transformations (parameters $\sqrt{1 + \partial_{\pm\pm}a^{\pm\pm}}$)
of the fermionic coordinates $\eta^{\pm}$.
Note, that
the first multiplier in the decomposition \p{suctr} is easily recognized as
the representatives of the left coset space $G/G_0$
\begin{eqnarray}\label{coset}
K^{\pm\pm}(\zeta) &=&
\xi^{\pm\pm} + i\eta^{\pm}\epsilon^{\pm}(\xi^{\pm\pm}), \\
K^{\pm}(\zeta) &=& \epsilon^{\pm}(\xi^{\pm\pm}) + \eta^{\pm}
\sqrt{1 + i\epsilon^{\pm}\partial_{\pm\pm}\epsilon^{\pm}}.
\nonumber
\end{eqnarray}
It deserves to mention that in the decomposition \p{suctr} the
comultipliers $K$ and $G_0$ are chosen in such a way that the irreduciblity
constraint \p{Del} is satisfied separately for each of them.

The prescription of constructing the corresponding nonlinear realization
proposed in \cite{ik} is as follows. Let us identify the local parameters
$\epsilon^{\pm}(\xi^{\pm\pm})$ in \p{coset} with the Goldstone fields
$\lambda^{\pm}(\xi^{\pm\pm})$
\begin{eqnarray}\label{cosetNR}
\tilde{K}^{\pm\pm}(\tilde{\zeta}) &=& \tilde{\xi}^{\pm\pm} +
i\tilde{\eta}^{\pm}\lambda^{\pm}(\tilde{\xi}^{\pm\pm}), \\
\tilde{K}^{\pm}(\tilde{\zeta}) &=& \lambda^{\pm}(\tilde{\xi}^{\pm\pm}) +
\tilde{\eta}^{\pm}
\sqrt{1 + i\lambda^{\pm}\tilde{\partial}_{\pm\pm}\lambda^{\pm}}.
\nonumber
\end{eqnarray}
and take for $\tilde{K}(\tilde{\zeta})$ the transformation law associated to
\p{suctr}
\begin{equation}\label{NR}
G(\tilde{K}(\tilde{\zeta})) = \tilde{K}^{\prime}(\tilde{G}_0(\tilde{\zeta})).
\end{equation}
In Eq. \p{NR} the newly introduced coordinates
$\tilde{\zeta} = \{\tilde{\xi}^{\pm\pm}, \tilde{\eta}^{\pm}\}$ are
transformed differently as
compared with $\zeta = \{\xi^{\pm\pm}, \eta^{\pm}\}$ in \p{G}. Indeed, in
accordance with \p{Gvac} they change only under the vacuum stability subgroup
\begin{eqnarray}\label{GvacNR}
\tilde{\xi}^{\pm\pm\prime} &=& \tilde{\xi}^{\pm\pm} +
\tilde{a}^{\pm\pm}(\tilde{\xi}^{\pm\pm}), \\
\tilde{\eta}^{\pm\prime} &=& \tilde{\eta}^{\pm}\sqrt{1 +
\tilde{\partial}_{\pm\pm}\tilde{a}^{\pm\pm}}.
\nonumber
\end{eqnarray}
where the parameters $\tilde{a}^{\pm\pm}(\tilde{\xi}^{\pm\pm})$ turn out to
be
dependent nonlinearly on the Goldstone fields $\lambda^{\pm}(\xi^{\pm\pm})$
and its derivatives.
Eqs. \p{NR} and \p{GvacNR}
determine the
transformation properties of the Goldstone fermions
$\lambda^{\pm}(\xi^{\pm\pm})$ with
respect to the nonlinear realization of the superconformal group $G$ in coset
space \p{cosetNR}.

\section{Splitting superfields and gauge relaxing}

Up to now we have dealt with only formal prescription of construction of the
nonlinear realization of superconformal group $G$ without any relation
of this procedure to the original equations \p{n11sl}, \p{gfe}.
Nevertheless, there is
the simple possibility to gain a more deeper insight into the model we
started with if we compare two Eqs. \p{G} and \p{NR}. We find that
$\tilde{K}(\tilde{\zeta})$ transform under $G$ in precisely the same manner
as the
initial coordinates $\zeta$ of superspace ${\bf R}^{(2 \mid 2)}$. Thus we
have the unique possibility to identify them
\begin{equation}\label{CV}
\zeta = \tilde{K}(\tilde{\zeta}).
\end{equation}
Eq. \p{CV} establishes the relationship between two forms of the realization
of superconformal symmetries in superspace, i.e. linear and nonlinear one.
One of the remarkable futures of the
transformations \p{CV} is that superspace of the nonlinear realization
${\bf \tilde{R}}^{(2 \mid 2)} = \{\tilde{\zeta}\}$ turns out to be completely
"splitting" in virtue of the transformations \p{GvacNR} which are not mixed
the bosonic and fermionic variables. Due to this very important fact the SFs
of the nonlinear realization valued in ${\bf \tilde{R}}^{(2 \mid 2)}$ becomes
{\it reducible}.
Furthermore we receive the unique possibility of relaxing the gauge
fixing Eqs. \p{gfe} because it appears that in frame of the nonlinear
realization there exist the new covariant objects consisting
only on the Goldstone fields which
transformed under the superconformal symmetry as the combinations of SFs
standing in the l.h.s. of these equations.
Indeed, let us consider the following quantities
\begin{eqnarray}\label{Phi}
\Phi^+_L(\xi^{++}, \eta^+) &\equiv&
\tilde{F}_L(\tilde{\xi}^{++})(\tilde{D}_+\eta^+)^{-3/2}, \\
\Phi^-_R(\xi^{--}, \eta^-) &\equiv&
\tilde{F}_R(\tilde{\xi}^{--})(\tilde{D}_-\eta^-)^{-3/2}.
\nonumber
\end{eqnarray}
Immediately from the definitions
and the connections
between the spinor covariant derivatives of linear and nonlinear realizations
$D_{\pm} = (\tilde{D}_{\pm}\eta^{\pm})^{-1}\tilde{D}_{\pm}$ one
can check
that these
objects are transformed as a superconformal densities of the weight $-3/2$
with respect to the superconformal transformations \p{sct}
\begin{eqnarray}\label{tpPhi}
\Phi^{+\prime}_L(\xi^{++\prime}, \eta^{+\prime}) &=&
(D_+\eta^{+\prime})^{-3/2}\Phi^+_L(\xi^{++}, \eta^+), \\
\Phi^{-\prime}_R(\xi^{--\prime}, \eta^{-\prime}) &=&
(D_-\eta^{-\prime})^{-3/2}\Phi^-_R(\xi^{--}, \eta^-),\nonumber
\end{eqnarray}
if the fields of nonlinear realization
$\tilde{F}_L(\tilde{\xi}^{++})$, $\tilde{F}_R(\tilde{\xi}^{--})$ are
supposed to be transformed as the corresponding densities with the same
weight with respect to ordinary conformal transformations
\begin{equation}\label{FLR}
\tilde{F}^{\prime}_L(\tilde{\xi}^{++\prime}) =
\omega_L^{-3/2}(\tilde{\xi}^{++})\tilde{F}_L(\tilde{\xi}^{++}),~~~
\tilde{F}^{\prime}_R(\tilde{\xi}^{--\prime}) =
\omega_R^{-3/2}(\tilde{\xi}^{--})\tilde{F}_R(\tilde{\xi}^{--}),
\end{equation}
where
\begin{equation}\label{omLR}
\omega_L \equiv \sqrt{1 + \partial_{++}a^{++}},~~~~
\omega_R \equiv \sqrt{1 + \partial_{--}a^{--}}.
\end{equation}
Therefore if we change the units in the r.h.s. of the Eqs. \p{gfe} on the SFs
\p{Phi} we obtain the equations
\begin{eqnarray}\label{ngfe}
D_+\Psi^+_L - 2(D_+W)\Psi^+_L = \Phi^+_L(\xi^{++}, \eta^+), \\
D_-\Psi^-_R - 2(D_-W)\Psi^-_R = \Phi^-_R(\xi^{--}, \eta^-), \nonumber
\end{eqnarray}
which will not restrict the gauge parameters.
Now let us return to the Eq. \p{n11sl}. Performing the change of variables
\p{CV} in the Eqs. \p{n11sl}, \p{ngfe} we get
\begin{equation}\label{tn11sl}
\tilde{D}_-\tilde{D}_+\tilde{W} =
e^{2\tilde{W}}\tilde{\Psi}^+_L\tilde{\Psi}^-_R,~~~
\tilde{D}_{\pm} = \tilde{\partial}_{\pm} +
i\tilde{\eta}^{\pm}\tilde{\partial}_{\pm\pm},
\end{equation}
\begin{eqnarray}\label{tgfe}
\tilde{D}_+\tilde{\Psi}^+_L - 2(\tilde{D}_+\tilde{W})\tilde{\Psi}^+_L
&=& \tilde{F}_L, \\
\tilde{D}_-\tilde{\Psi}^-_R - 2(\tilde{D}_-\tilde{W})\tilde{\Psi}^-_R
&=& \tilde{F}_R, \nonumber
\end{eqnarray}
where the SFs and covariant
derivatives of the nonlinear realization \p{NR}, \p{GvacNR} and \p{CV}
are introduced
\begin{eqnarray}\label{NRSF}
W(\xi^{\pm\pm}, \eta^{\pm}) &=&
\tilde{W}(\tilde{\xi}^{\pm\pm}, \tilde{\eta}^{\pm})
 - \frac{1}{4}ln(\tilde{D}_+\eta^+) -
\frac{1}{4}ln(\tilde{D}_-\eta^-), \\
\Psi^+_L(\xi^{++}, \eta^+) &=&
(\tilde{D}_+\eta^+)^{-1/2}\tilde{\Psi}^+_L(\tilde{\xi}^{++}, \tilde{\eta}^+), \nn
\Psi^-_R(\xi^{--}, \eta^-) &=&
(\tilde{D}_-\eta^-)^{-1/2}\tilde{\Psi}^-_R(\tilde{\xi}^{--}, \tilde{\eta}^-).
\nonumber
\end{eqnarray}
Note that although the form of the Eq. \p{tn11sl} is precisely the same as
the
original one \p{n11sl} the SFs of the nonlinear realization appearing in
\p{tn11sl}, \p{tgfe} are distinguished drastically from the SFs of linear
realization. As it follows from \p{GvacNR} the "new" SFs $\tilde{W}$ and
$\tilde{\Psi}$
are transformed under the action of $G$ only with respect to their stability
subgroup \p{GvacNR}
\begin{eqnarray}\label{gttSF}
\tilde{W}^{\prime}(\tilde{\xi}^{\pm\pm\prime}, \eta^{\pm\prime}) &=&
\tilde{W}(\tilde{xi}^{\pm\pm}, \tilde{\eta}^{\pm})
 - \frac{1}{4}ln(\tilde{D}_+\tilde{\eta}^{+\prime}) -
\frac{1}{4}ln(\tilde{D}_-\tilde{\eta}^{-\prime}), \\
\tilde{\Psi}^{+\prime}_L(\tilde{\xi}^{++\prime}, \tilde{\eta}^{+\prime}) &=&
(\tilde{D}_+\tilde{\eta}^{+\prime})^{-1/2}
\tilde{\Psi}^+_L(\tilde{\xi}^{++}, \tilde{\eta}^+), \nn
\tilde{\Psi}^{-\prime}_R(\tilde{\xi}^{--\prime}, \tilde{\eta}^{-\prime}) &=&
(\tilde{D}_-\tilde{\eta}^{-\prime})^{-1/2}
\tilde{\Psi}^-_R(\tilde{\xi}^{--}, \tilde{\eta}^-).
\nonumber
\end{eqnarray}
Substituting here the explicit form of gauge parameters deduced from the
transformations \p{GvacNR}
\begin{equation}\label{Deta}
\tilde{D}_{\pm}\tilde{\eta}^{\pm\prime} =
\sqrt{1 + \tilde{\partial}_{\pm\pm}\tilde{a}^{\pm\pm}},
\end{equation}
one concludes that all the component fields of the SFs $\tilde{W}$ and
$\tilde{\Psi}^{\pm}$ are transformed {\it independently} from each other.
Thus we can put down the following manifestly covariant constraints
\begin{equation}\label{Wconstr}
\tilde{W}(\tilde{\xi}^{\pm\pm}, \tilde{\eta}^{\pm}) =
\tilde{u}(\tilde{\xi}^{\pm\pm}),
\end{equation}
\begin{equation}\label{Pconstr}
\tilde{\Psi}^+_L = \tilde{\eta}^+\tilde{F}_L~~\Rightarrow~~\tilde{\omega}^+
= 0,~~
\tilde{\Psi}^-_R = \tilde{\eta}^-\tilde{F}_R~~\Rightarrow~~\tilde{\omega}^-
= 0.
\end{equation}
It is instructive to note that the two last constraints in \p{Pconstr} are
specific for the theories with spontaneously broken local symmetries. They
establish the equivalence connections between the Goldstone fields of linear
and nonlinear realizations. In the case under consideration one can proves
that this connection between the corresponding fields $\omega^{\pm}$ and
$\lambda^{\pm}$ arise only when the component
fields $F_{L,R}$ are developed the nonzero vacuum expectation values. Another
fact which is more suggestive in our opinion is that the gauge freedom of
the residual system remained in our disposal allows one to put the gauge in
which $\lambda^{\pm} = 0$. This is well-known unitary gauge which always can
be achieved in any theory with the spontaneously broken gauge symmetry.

Returning the SFs \p{Wconstr}, \p{Pconstr} back into the system \p{tn11sl},
\p{tgfe} we obtain the ordinary Liouville equation {\it only}
\begin{equation}\label{LEq}
\tilde{\partial}_{--}\tilde{\partial}_{++}\tilde{u} =
e^{2\tilde{u}}\tilde{F}_L\tilde{F}_R.
\end{equation}
Two remainder Eqs. \p{tgfe} are satisfied identically due to the
constraints \p{Pconstr}.

\section{General solution}
Let us consider shortly the problem of construction of general solution
of the Eqs. \p{n11sl}, \p{gfe}. It is obvious that this solution can be
obtained directly from the residual Eq. \p{LEq}. Indeed, we know from
\cite{biku} that the corresponding solution can be written in form
\begin{eqnarray}\label{LEqsol}
e^{-2\tilde{u}(\tilde{\xi}^{\pm\pm})} &=&
\frac{1}{2}\tilde{r}_m^{++}(\tilde{\xi}^{--})
\tilde{l}^{--m}(\tilde{\xi}^{++}), \\
\tilde{F}_L(\tilde{\xi}^{++}) &=&
\tilde{l}_m^{--}(\tilde{\xi}^{++})\tilde{\partial}_{++}
\tilde{l}^m(\tilde{\xi}^{++}), \nn
\tilde{F}_R(\tilde{\xi}^{--}) &=&
\tilde{r}_m^{++}(\tilde{\xi}^{--})
\tilde{\partial}_{--}\tilde{r}^m(\tilde{\xi}^{--}),
\nonumber
\end{eqnarray}
where the left(right)-moving Lorentz harmonics are normalized as follows
\begin{equation}\label{orthog}
\tilde{l}_m^{++} \tilde{l}^{m++} = 0,\quad
\tilde{l}_m^{--} \tilde{l}^{m--} = 0,\quad
\tilde{l}_m
\tilde{l}^{m\pm\pm} = 0,
\end{equation}
\begin{equation}\label{norm}
\tilde{l}_{m}^{--}
\tilde{l}^{m++} = 2,\quad
\tilde{l}_m
\tilde{l}^m = -1.
\end{equation}
Performing here the change of the variables inverse relative to \p{CV} one
can always reaches the general solution of the Eqs. \p{n11sl} and \p{ngfe}
in terms of harmonic SFs restricted by the constraints
\begin{eqnarray}\label{LRsol}
l_m^{\pm\pm,0}(\xi^{++}, \eta^+) &=&
\tilde{l}_m^{\pm\pm,0}(\tilde{\xi}^{++}), \\
r_m^{\pm\pm,0}(\xi^{--}, \eta^-) &=&
\tilde{l}_m^{\pm\pm,0}(\tilde{\xi}^{--}).
\nonumber
\end{eqnarray}

\section{Conclusion}
Thus we have demonstrated that owing to
the relaxation of the gauge fixing auxiliary equation of motion \p{gfe}
$\Rightarrow$ \p{ngfe} we obtain the ordinary Liouville equation \p{LEq}
instead of SF Eq. \p{n11sl}. This is very important result because it clarify
the general method of construction of the superstring-inspired nonlinear
equations of motion in the case of arbitrary space-time dimension $D$
starting directly from the corresponding equation of motion in the bosonic
sector. Indeed, let us suppose that we have know the linear realization of
the supergroup $G$ in the worldsheet superspace $n=(p,p),~~p=1,2,4,8$, which
describes the corresponding superconformal transformations. Decomposing of an
arbitrary element of this group onto the product of two elements, i.e.
proper chosen coset
space and that of the stability subgroup $G=KG_0$, we can always obtain the
suitable nonlinear realization of $n=(p,p)$ superconformal symmetry following
closely to the pattern of the Section 2. Then the sought for form of the SF
equation of motion could be deduced with the help of the transformations
which are inverse relative to $\zeta = \tilde{K}(\tilde{\zeta})$.

\section*{Acknowledgments}

It is a great pleasure for me to express grateful to E.~Ivanov,
S.~Krivonos, and A.~Pashnev for interest to this work
and valuable discussions. I would like also to thank Prof. A. Filippov for
kind invitation and hospitality at the Laboratory of Theoretical Physics,
where this work was done.

\end{document}